\documentclass[twocolumn]{aastex63}

\received{2021 August 25}
\revised{2022 July 31}
\accepted{2022 August 2}
\submitjournal{ApJ}

\shorttitle{MIR and maser variability in G036.70+00.09}
\shortauthors{Uchiyama et al.}

\usepackage{amsmath}
\usepackage{color}
\usepackage{url}
\usepackage{ulem}
\usepackage{multirow}
\usepackage{longtable}
\usepackage{natbib}
\newcommand{\wise}{\textit{WISE}}
\newcommand{\allwise}{\textit{ALLWISE}}

\usepackage{graphicx}
\usepackage{epstopdf}

\begin{document}

\title{Mid-Infrared and Maser Flux Variability Correlation in Massive Young Stellar Object G036.70+00.09}

\correspondingauthor{Mizuho Uchiyama}
\email{uchiyama@ir.isas.jaxa.jp, k.ichikawa@astr.tohoku.ac.jp}

\author[0000-0002-6681-7318]{Mizuho Uchiyama}
\affil{Institute of Space and Astronautical Science, Japan Aerospace Exploration Agency, 3-1-1 Yoshinodai, Chuo-ku, Sagamihara 252-5210, Japan}

\author[0000-0002-4377-903X]{Kohei Ichikawa}
\affil{Max-Planck-Institut f{\"u}r extraterrestrische Physik (MPE), Giessenbachstrasse 1, D-85748 Garching bei M{\"u}unchen, Germany
}
\affil{Frontier Research Institute for Interdisciplinary Sciences, Tohoku University, Sendai 980-8578, Japan}
\affil{Astronomical Institute, Graduate School of Science
Tohoku University, 6-3 Aramaki, Aoba-ku, Sendai 980-8578, Japan}

\author[0000-0002-6033-5000]{Koichiro Sugiyama}
\affil{National Astronomical Research Institute of Thailand (Public Organization), 260 Moo 4, T. Donkaew, A. Maerim, 
Chiangmai, 50180, Thailand}
\affil{Mizusawa VLBI Observatory, National Astronomical Observatory of Japan, 2-21-1 Osawa, Mitaka, Tokyo 181-8588, Japan}

\author{Yoshihiro Tanabe}
\affil{Center for Astronomy, Ibaraki University, 2-1-1 Bunkyo, Mito, Ibaraki 310-8512, Japan}

\author[0000-0001-5615-5464]{Yoshinori Yonekura}
\affil{Center for Astronomy, Ibaraki University, 2-1-1 Bunkyo, Mito, Ibaraki 310-8512, Japan}

\begin{abstract}
We present the discovery of the simultaneous flux variation of a massive young stellar object (MYSO) G036.70+00.09 (G036.70) both in the maser emission and mid-infrared (MIR;$\lambda=3$--$5$~$\mu$m) bands.
Utilizing the ALLWISE and NEOWISE archival databases covering a long time span of approximately 10 years with a cadence of 6 months, we confirmed that G036.70 indicates a stochastic year-long MIR variability with no signs of the WISE band color change of W1 (3.4~$\mu$m) $-$W2 (4.6~$\mu$m).
Cross-matching the MIR data set with the high-cadence 6.7~GHz class II methanol maser flux using a Hitachi 32-m radio telescope that discovered its periodicity in the methanol maser of 53.0--53.2 days, we also determine the flux correlations between the two bands at two different timescales, year-long and day-long, both of which have never been reported in MYSOs except when they are in a state of the accretion burst phase.
The results of our study support the scenario that a class II methanol maser is pumped up by infrared emission from accreting disks of MYSOs.
We also discuss the possible origins of MIR and maser variability.
To explain the two observed phenomena, a stochastic year-long MIR variability with no signs of significant color change and maser-MIR variability correlation, change in mass accretion rate and line-of-sight extinction because of nonaxisymmetric dust density distribution in a rotating accretion disk are possible origins.
Observations through spectroscopic monitoring of accretion-related emission lines are essential for determining the origin of the observed variability in G036.70.
\end{abstract}


\section{Introduction}

Massive stars play a significant role in star formation activities and metal enrichment in galaxies and, hence, in the evolution of galaxies \citep[for example,][and references therein]{ZY07,Tan14} 
as well as growth of the supermassive black holes in the universe \citep[for example,][]{tho05,ina16a,ich17b}.
However, owing to observational difficulties, the detailed views of the youngest phase of massive stars, massive young stellar objects (MYSOs), have yet to be resolved.

Variability studies are a promising means to investigate distant and compact objects such as mass-accreting MYSOs and their disks and/or outflows.
For lower-mass YSOs, owing to the small extinction of dust with $A_V < 10$ \citep{MB99}, observations in both the optical and near-infrared (NIR; $\lambda<3$~$\mu$m) bands have been well conducted over the last 50 years. 
Such optical and NIR variability studies involving the emission lines and continuum of lower-mass YSOs have revealed the physics of star formation and pre-main-sequence stellar evolution, such as the existence of rotating cool/hot spots \citep{BB89}, infalling/rotating dusty objects \citep{Cody14}, and drastic mass-accretion rate variation \citep{Audard14}.
However, for MYSOs, such optical or NIR systematic variability surveys were not reported until recently, due to huge extinction towards MYSOs and hence difficulty of optical and NIR observations.

Recently, high-amplitude ($\Delta K_\mathrm{s} > 1$~mag) year-scale variability in MYSOs was reported for the first time in the $K_\mathrm{s}$ band using the Vista Variables in Via Lactea (VVV) survey data \citep{Kumar16}.
Following that study, lower-amplitude variables, approaching $\Delta K_\mathrm{s} > 0.15$~mag, were detected in 190 out of 718 MYSO candidates in the same VVV survey data \citep{Teixeira18}.
This discovery motivates authors to search for variability at longer wavelengths.
Considering that MYSOs are heavily obscured by their surrounding gas and dust, searching for variability in the mid-IR (MIR; $\lambda>3~\mu$m) is more suitable owing to its relative insensitivity towards dust extinction.  \citet{uch19} recently reported $\sim1$~yr scale MIR variable MYSOs by utilizing the combined ALLWISE and NEOWISE data covering over an $\sim10$~yr observation period with 6~month cadence \citep{wri10,mai11,mai14}.

Radio bands also provide significant information mostly from non-thermal emissions for certain MYSOs. These are observed as masers, which often indicate periodic and/or non-periodic flux variations in the radio bands. Some masers are classified as strongly related to MYSOs \citep[for example,][]{Minier03,Breen13,Motogi16}.
In particular, certain class II methanol masers are considered to be associated with accreting disks of MYSOs \citep[e.g.,][]{Sugiyama14,Motogi17,Sanna17} with a typical size of several hundreds of AU \citep{Tan14}, pumped up by thermal radiation in the MIR \citep{Cragg05}.
The origins of periodic maser activity have been widely discussed in the previous decade, such as the periodic accretion of circumbinary material in binary system \citep{Araya10}, colliding wind binary \citep{vanderWalt09,deWalt11}, stellar pulsation of an accreting MYSO \citep{Inayoshi13}, rotation of spiral shocks in the gaps of circumbinary disks \citep{Parfenov14}, and a low-mass YSO blocking ultraviolet radiation from a high-mass star in an eclipsing binary system \citep{Maswanganye15}.
Recently, flux correlation between a 6.7~GHz methanol maser and infrared emission has been reported in a drastically variable burst MYSO \citep{CoG17, uch20,Stecklum21}.
For periodic variability, such a correlation is reported in the intermediate-mass YSO, G107.298+5.639 \citep{Olech20}.
However, a study on the variability correlations between infrared and masers in MYSOs associated with periodic 6.7 GHz maser as well as IR color variability during methanol maser flux variation is still unavailable.

In this study, we investigate the variability and the flux variability correlations of the MIR and 6.7~GHz methanol maser emissions for MYSO G036.70+00.09 (hereafter, referred to as G036.70) (the absolute coordinate from \citet{Bart09,Pandian11} in J2000.0: $R.A.=18:57:59.123$, Decl.$=+03:24:06.12$);
this methanol maser indicates a periodic flux variability with a period of $\sim$ 53 d \citep{Sugiyama2015,Sugiyama19}.
We use the methanol maser data observed in a long-term and high-cadence monitoring program conducted using the Hitachi 32-m radio telescope \citep{Yonekura16, Sugiyama19} and WISE archival MIR data, particularly at bands 1 (W1;~3.4~$\mu$m) and 2 (W2;~4.6~$\mu$m).

\section{Archival Data}

\subsection{Cross-matching with WISE}
We obtained the MIR counterparts of G036.70 through positional 
matching with \wise\ data to generate the MIR (3--5~$\mu$m) light curve.
We collected multi-epoch photometry data
from the \allwise\  \citep[for all four bands;][]{wri10} and
the most recent \textit{NEOWISE} \citep[only for W1 and W2; ][]{mai11,mai14}
2020 data release\footnote{\url{http://wise2.ipac.caltech.edu/docs/release/neowise/}} covering observations between
December 13, 2013 (MJD 56639) and December 13, 2019 (MJD 58830).
\textit{WISE} has a 90-min orbit and conducts $\approx 12$ observations
of a source over a $\approx 1$-d period and visits a specified location every six months.

In this study, we conducted a cross-matching procedure in the same manner as conducted by \cite{uch19}.
We summarize the key steps here.
We applied a cross-matching radius of 2~arcsec based on the positional accuracy using the 2MASS catalog \citep[for example, see ][]{ich12,ich17a,ich19}.
These initial data points comprised 179 multi-epoch photometric data points.

\begin{figure*}
\begin{center}
\includegraphics[width=0.45\linewidth]{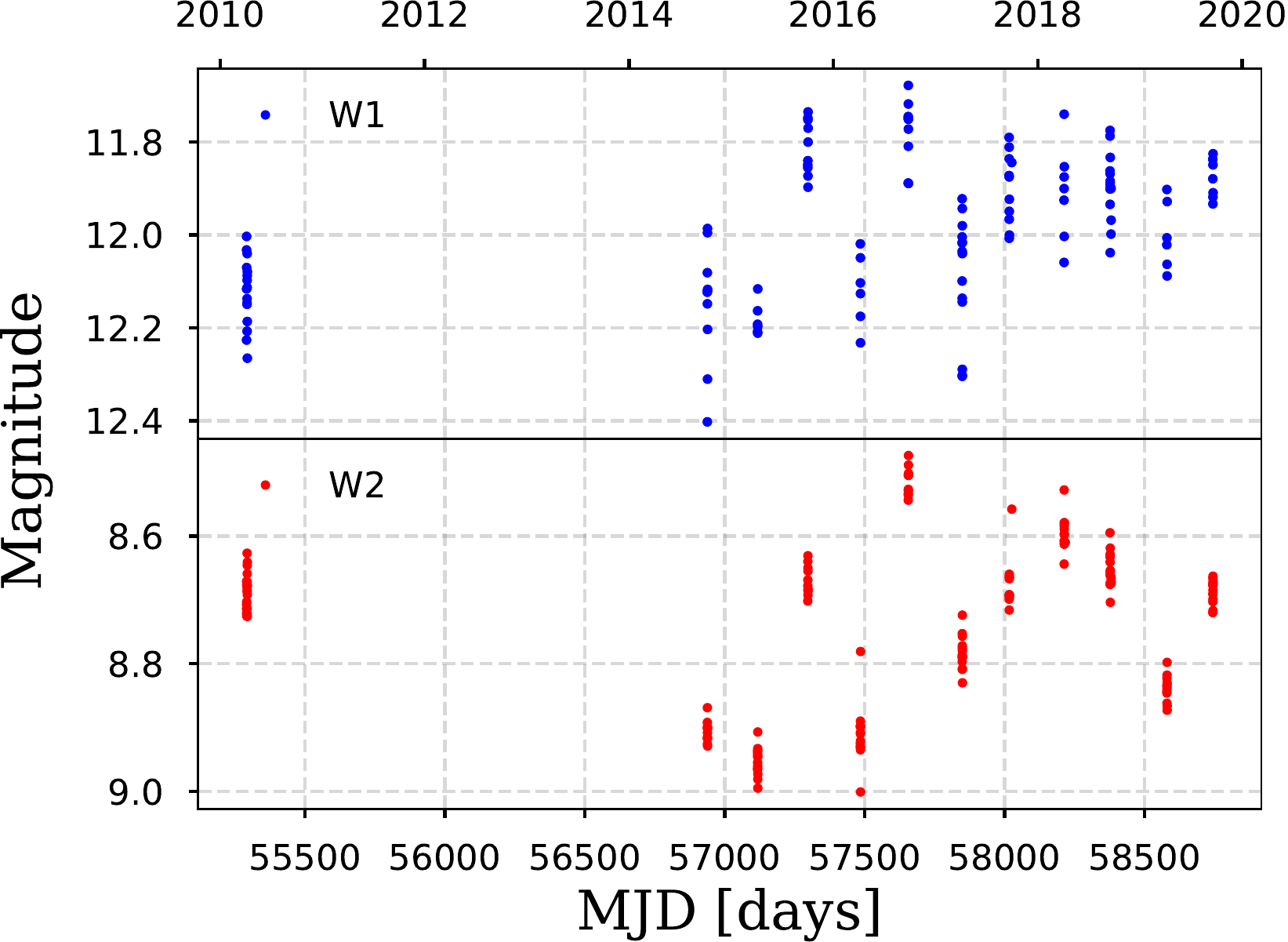}~
\includegraphics[width=0.55\linewidth]{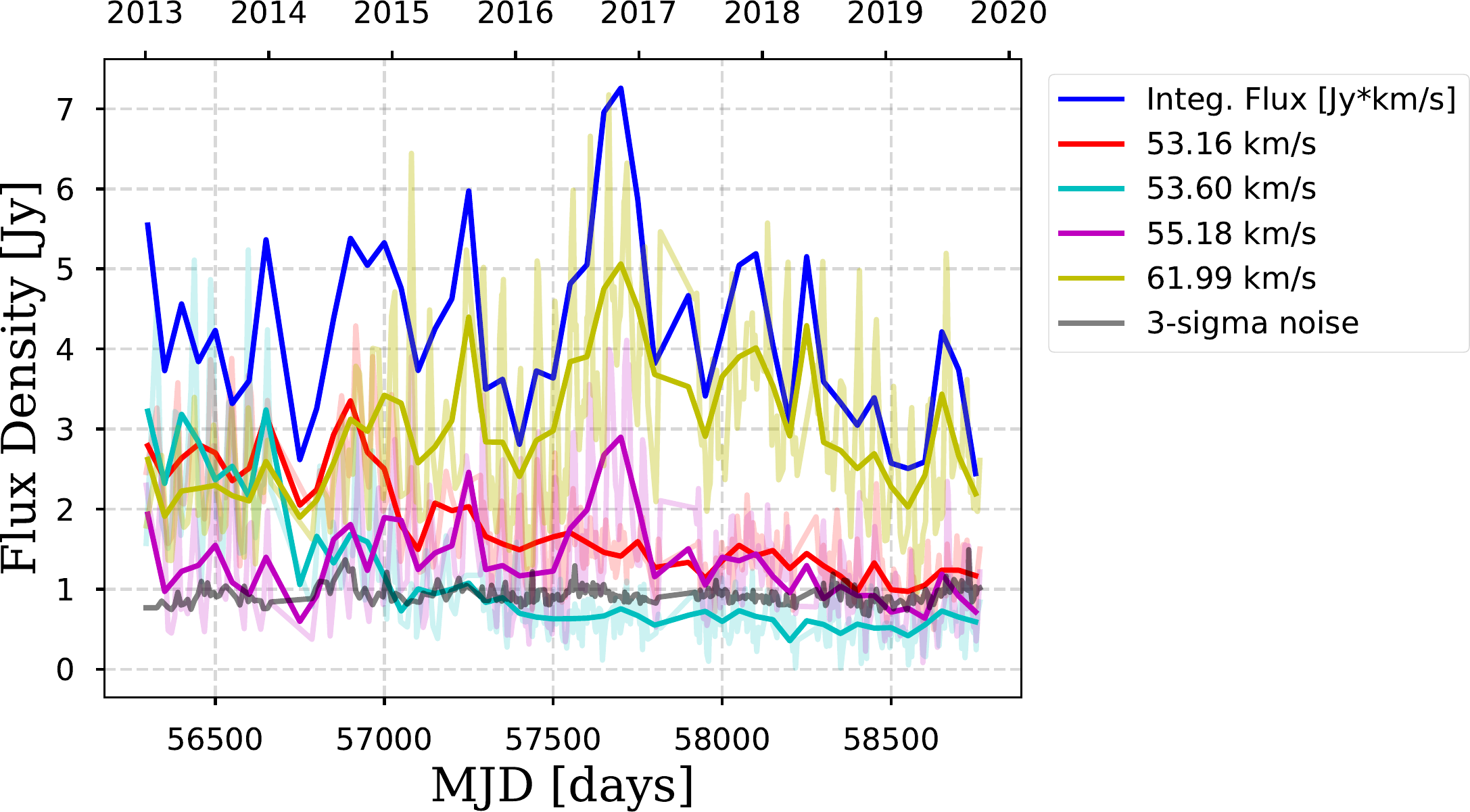}
\caption{
\textit{Left}: MIR light curves of G036.70 in the WISE W1 (blue; top panel) and W2 (red; bottom panel) data.
 The single selected exposures are presented with the error values.
\textit{Right}: Light curves of G036.70 in 6.7-GHz methanol maser data. Each color represents each major velocity component of the 6.7-GHz methanol maser. The faint lines represent the original data, whereas the thick lines represent the 50-day binning data to observe year-scale variability. 
The thick blue line represents the velocity integrated flux with the unit of Jy~km~s$^{-1}$. 
The data around MJD$\sim 56700$ and MJD$\sim57800$ are missing owing to the maintenance of the facility.} 
\label{fig:LC}
\end{center}
\end{figure*}

We then applied additional parameters to obtain reliable photometric data.
We used a profile-fitting magnitude \texttt{w1/2mpro} with
reliable photometric qualities, fulfilling a signal-to-noise
ratio larger than 10.0 (\texttt{ph\_qual=A})
and having the best image quality (\texttt{qi\_fact$\geq$0.5}).
We further selected the photometric data not affected by artificial source contamination, South Atlantic Anomaly, or moon-light, \texttt{qual\_frame>0.0}, \texttt{saa\_sep>5.0}, and \texttt{moon\_masked=0},
as well as the contamination flag as \texttt{cc\_flags=0},
which are known to be unaffected by the known artifacts.
MYSOs are often very bright, reaching the saturation limit at $\sim$8
and 7 mags in W1 and W2 bands. To avoid such unreliable
photometric data, we selected photometric points with less than 5\% (\texttt{w1/2\_sat<0.05}) saturated pixels
and then removed the data with failed background sky fitting
and bad point spread function (PSF) profile fitting, flagged as \texttt{w1/2\_sky=NAN} and \texttt{w1/2\_rchi2$\leq$150}.
This selection process finally retains 116 and 160 photometric data points for G036.70 in W1 and W2 bands, respectively,
as depicted in Figure~\ref{fig:LC}.

\subsection{6.7-GHz Methanol Maser Data}

To compare with WISE archival data, we used a part of the 6.7 GHz methanol maser monitoring data observed by the Hitachi 32-m radio telescope \citep{Yonekura16,Sugiyama2015,Sugiyama19}. 
The velocity coverage of the data is $\sim$ 360 km/s, with the velocity resolution of $\sim$ 0.044 km/s. The typical RMS noise level (1$\sigma$) was $\sim$ 0.3 Jy.
The traditional chopper-wheel method \citep{UH1976} was used for the absolute flux calibration, yielding a typical 1$\sigma$ error of 7\%.
The flux decrease due to the pointing error is estimated to be as much as $\sim$ 3\%, which can be calculated from the beam size at 6.7 GHz of $\sim$4.6 arcmin and the typical pointing error of $<$ 30 arcsec. Thus, in total, the error of the observed flux density is estimated to be $<$ 10\%.
Between December 2012 and August 2015, monitoring observations of the intervals of 10 d were conducted. More intense monitoring of the intervals of 4--5 d was conducted for G036.70 after September 2015 until now. 
Here, we utilize the observation data obtained from January 2013 until October 2019 (MJD$=56295$--$58784$). 
Therefore, these methanol maser monitoring data associated with MYSOs are suitable for comparison with WISE archival one to study the correlation of variability between them.

\section{Results}\label{sec:results}

\subsection{Overview of Light Curves in the MIR and 6.7~GHz Maser Emission}

Figure~\ref{fig:LC} presents the light curves of WISE MIR bands (left panel; W1/W2 in the top/bottom) and the 6.7~GHz maser emission (right panel).
In the right panel plot, the thick lines represent the 50-day binning radio data to observe year-scale variability to compare with WISE MIR data easily.
Both the MIR and radio wavelengths indicate year-scale variability. 
The magnitudes of flux variation 
span 0.4~mag both in the W1 and W2 bands, whereas radio observations indicate as much as a $\sim$ 4 times 
magnification of flux 
(see the purple and yellow lines).
The radio light curves also indicate shorter time scale variabilities, which are prominent as several spike features in the right panel of Figure~\ref{fig:LC}. 
In contrast, MIR indicates
only the stochastic flux variation in the averaged WISE data, although it does not have a time resolution of less than a half year, and periodic variability less than that period cannot be detected in these data.

\subsection{Periodic Flux Variation of the 6.7 GHz Methanol Maser Emission in G036.70}

\begin{figure*}[]
\begin{center}
\includegraphics[width=\linewidth]{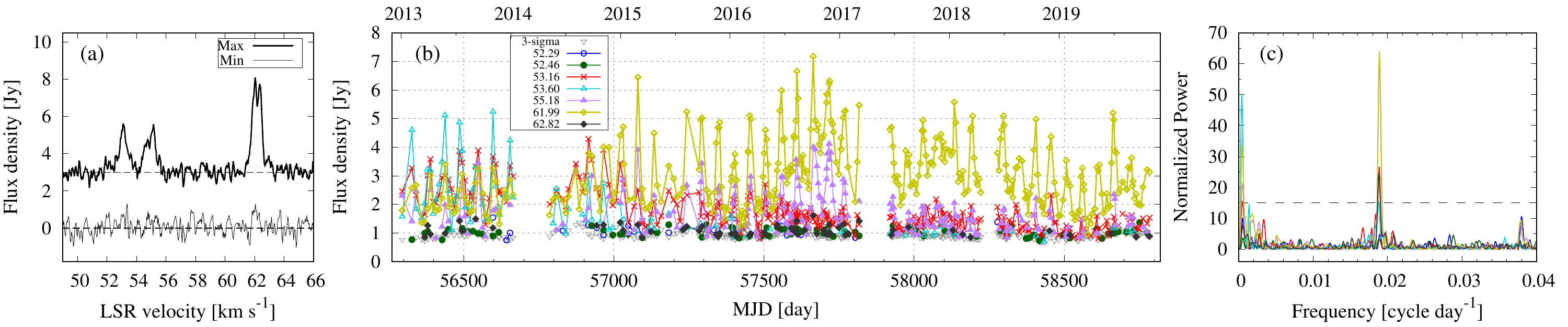}
\renewcommand{\baselinestretch}{0.7}%
\caption{
The 6.7~GHz methanol maser in G036.70 monitored with the Hitachi 32-m radio telescope, which was compiled in the term from January 3, 2013 to October 28, 2019 (MJD 56295--58784). 
(a) Spectrum for showing the maximum and minimum flux density at the peak spectral component at LSR velocity of $+$61.99~km~s$^{-1}$, respectively, in a term from February to April 2016 (MJD 57419--57479). The horizontal dashed lines correspond to baselines of 0~Jy in each spectrum. 
(b) Plot of flux variations for seven spectral components in the monitoring term, respectively (see the legend at the top-middle in the plot for symbols and colors showing each different component, in which their LSR velocities determined at the beginning of the monitoring). Inverted triangles with gray color correspond to the detection limit in each observation as three times the rms noise level. Top and bottom horizontal axis represents year and MJD, respectively. 
(c) Lomb-Scargle periodogram applied to all the spectral components. The definition of colors is the same to the panel (b). A false-alarm probability $\leq 10^{-4}$, corresponding to a normalized power level of 15.0, is denoted by a dashed horizontal line. }
\label{fig3-1}	
\end{center}	
\end{figure*}

The 6.7~GHz methanol maser in G036.70 was first detected through the unbiased survey with the Torun 32-m radio telescope \citep{2002A&A...392..277S}, in which showed seven spectral components in the LSR velocity range from $+ 52$ to $+ 63$~km~s$^{-1}$ \citep{2002A&A...392..277S,2007ApJ...656..255P,2015MNRAS.450.4109B}. This methanol maser has been monitored with the Hitachi 32-m radio telescope since January 3, 2013 (MJD 56295) as a target in the Ibaraki 6.7-GHz Methanol Maser Monitor (iMet) program \citep{Yonekura16}. 
We compiled the flux monitoring data until October 28, 2019 (MJD 58784) in this paper consisting of 328 observations data set, as shown in Figure~\ref{fig3-1}. During the monitoring term, there are lack of observations in some blank dates when the monitoring was not temporary performed due to system maintenance or other observations as follows: January 8 -- May 7, 2014 (MJD 56665--56784), March 6 -- June 14, 2017 (MJD 57818--57918), and April 15 -- June 7, 2018 (MJD 58223--58276). As an example of the spectrum, in a term from February to April 2016, the brightest spectral components at LSR velocity of $+ 61.99$~km~s$^{-1}$ (determined at the beginning of this monitoring) showed the minimum flux density of 1.26~Jy that is just equal to 4~$\sigma$ (: rms noise level) on February 4, 2016 (MJD 57422) and the maximum one of 5.10~Jy on March 5, 2016 (MJD 57452), respectively (Figure~\ref{fig3-1}a). At the maximum phase, other five spectral components at LSR velocities of $+ 52.46$, $+ 53.16$, $+ 53.60$, $+ 55.18$, and $+ 62.82$~km~s$^{-1}$ were brightened up beyond 3~$\sigma$. In other different phases, a remaining spectral component at $+ 52.29$~km~s$^{-1}$ occasionally came up in the spectra. 

These methanol maser emissions present a periodic flux variation with a continuous pattern, as shown in Figure~\ref{fig3-1}~(b). This periodicity was discovered through the iMet program, which was reported in \citet{Sugiyama2015,2018IAUS..336...45S,Sugiyama19}, via compiling the monitoring data until the mid-2016. We readopted the Lomb-Scargle (L-S) periodogram \citep{1976Ap&SS..39..447L,1982ApJ...263..835S} to update the period of each cycle in the flux variation until October 28, 2019 (MJD 58784). The L-S periodogram is the most reliable method to search for the periodicity in flux variations of methanol masers verified in \citet{2014MNRAS.437.1808G}. With an oversampling factor of four and a false-alarm probability $\leq 10^{-4}$ as significant that were used in the adoptions so far for other periodic sources \citep[e.g.,][]{2014MNRAS.437.1808G,Sugiyama17}, this readoption resulted in a period of 53.0~day (corresponding to a frequency of 0.01888~cycle~day$^{-1}$)  for the brightest spectral component at $+ 61.99$~km~s$^{-1}$ with a normalized power in the periodogram of 64.0 that was significantly beyond the power level of 15.0 corresponding the false-alarm probability of $10^{-4}$ (the dashed horizontal line), as shown in Figure~\ref{fig3-1}~(c). The L-S periodogram applied to other spectral components at $+ 53.16$, $+ 55.18$ and $+ 53.60$, $+ 62.82$~km~s$^{-1}$ also presented periods of 53.0 and 53.2~days, respectively, which were significantly detected beyond the false-alarm probability of $10^{-4}$. These periods were calculated from the neighboring frequency points on the L-S periodogram, respectively, in which the frequency resolution was 0.0001~cycle~day$^{-1}$ or $\sim$0.3~day. 

We summarize properties for the periodic flux variation of the 6.7~GHz methanol maser in G036.70 with the continuous pattern discovered through the iMet program as follows: i) its period of 53.0--53.2~days, ii) this periodicity was detected for most of the spectral components, iii) this continuous periodic flux variation has lasted in 47 cycles in the whole monitoring term of 2,489~days from January 3, 2013 to October 28, 2019 (MJD 56295--58784), although a part of these were not directly verified due to the lack of observations in the following durations: MJD 56665--56784, 57818--57918, and MJD 58223--58276.

\begin{figure*}
\begin{center}
\includegraphics[width=0.5\linewidth]{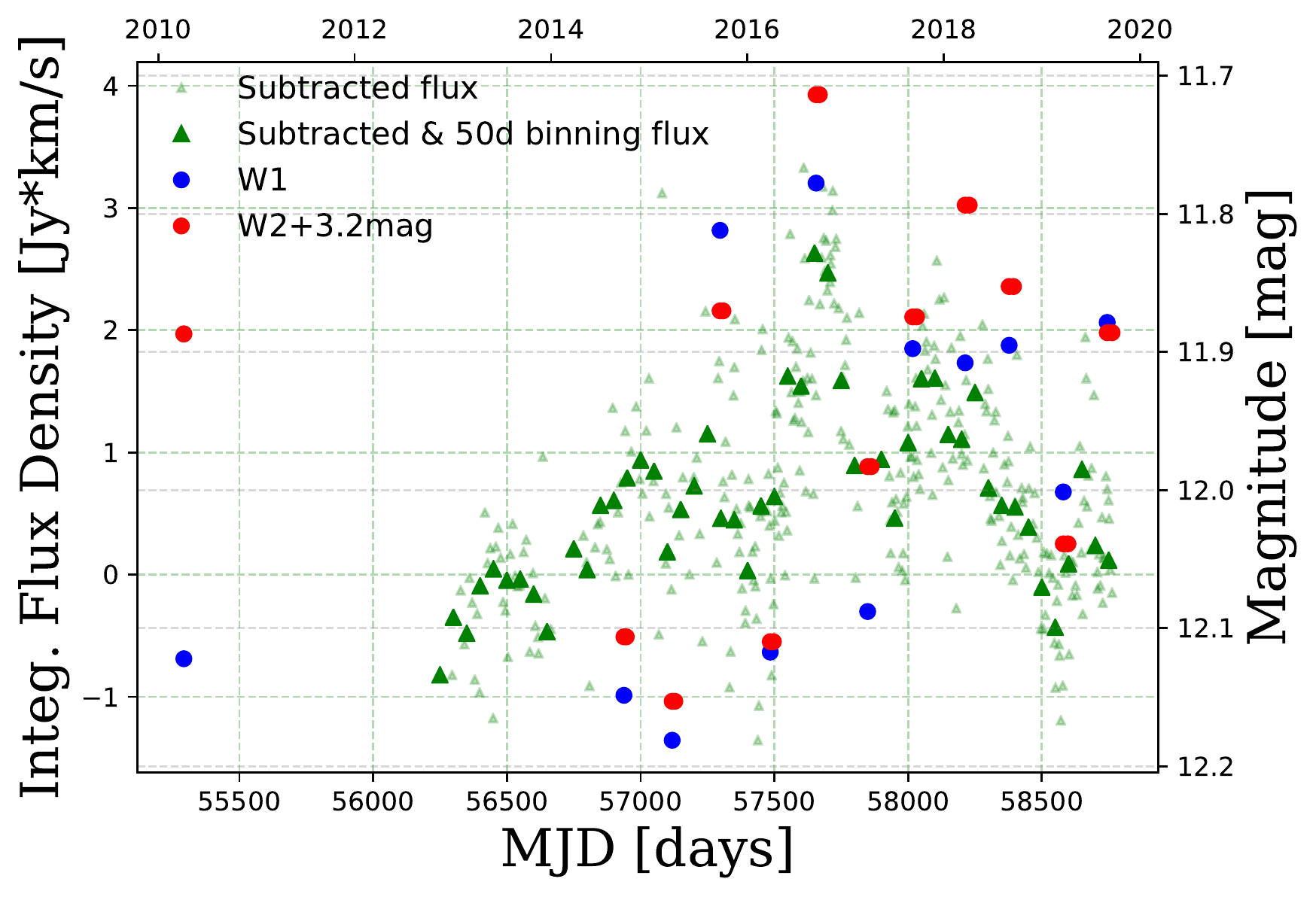}~
\includegraphics[width=0.5\linewidth]{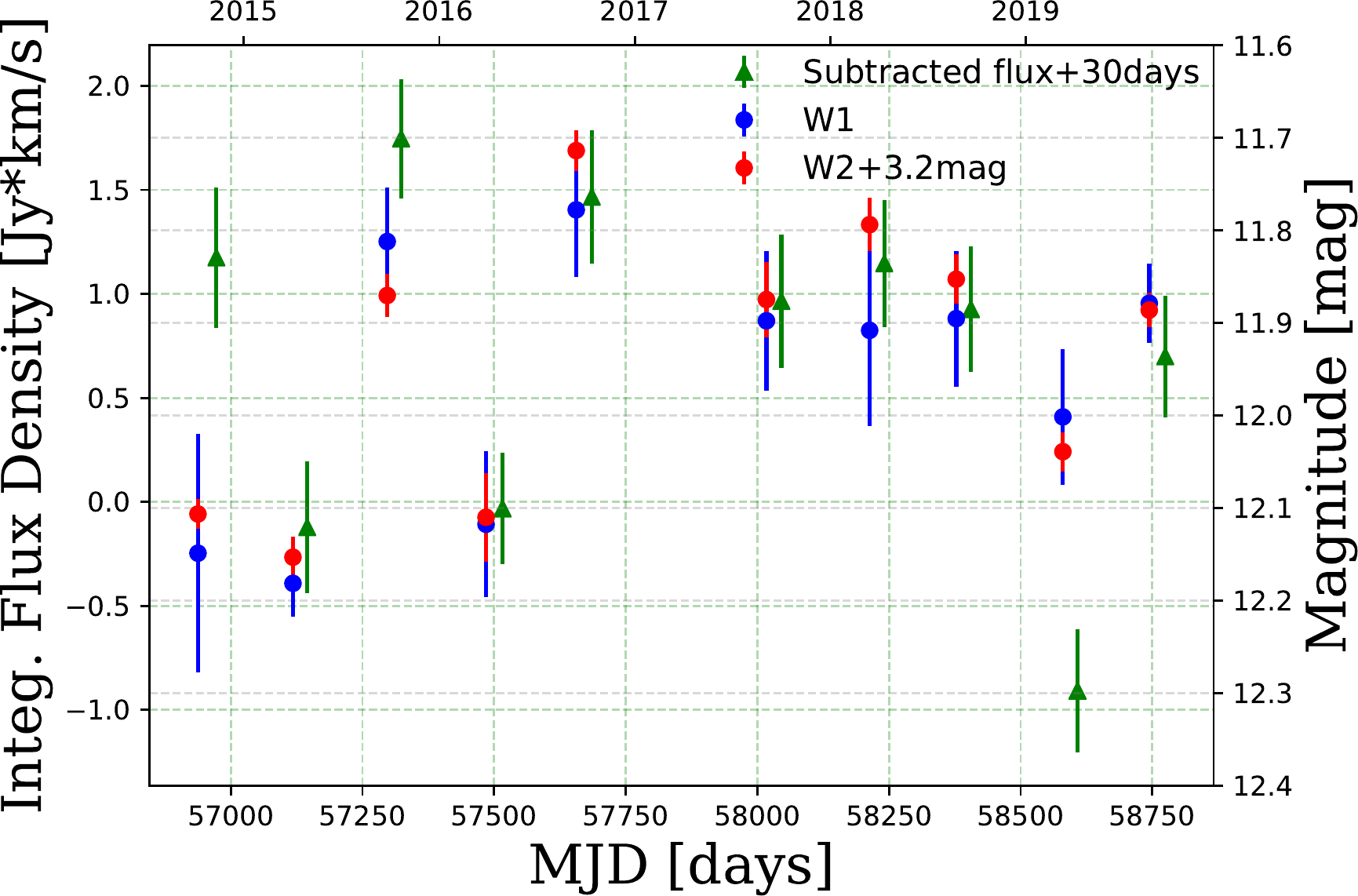}
\caption{
\textit{Left}: 
MIR light curves of G036.70 in the WISE W1 (blue circle) and W2 (red circle) data, together with maser fluxes (pale green triangle) after the flux subtraction of the known 53.0--53.2 day periodic variation. The dark green triangle represents the 50-d binning data of the maser light curves as discussed in Section~3.3.
\textit{Right}:Light curves of MIR data and ``matched'' maser data within 5 d or less. Maser data are shifted by 30 d to compare them clearly with the MIR data because data points of maser and MIR data apparently overlap without shifting. Note that this shifting is conducted only in this figure, not in data analysis.}
\label{fig:CMP}
\end{center}
\end{figure*}

\subsection{Comparison Between
MIR and 6.7~GHz Maser Light Curves}

To compare the MIR and radio light curves using different timescales,
two graphs were plotted as shown in Figure~\ref{fig:CMP}.
Hereafter, the integrated flux density is used as characteristic data of 6.7 GHz maser fluxes in comparison to MIR data.
First, the long, year-scale variability is compared in the left panel. 
In this plot, MIR single-exposure data are averaged over each scanning epoch for WISE, which is approximately 1-d duration and whose scanning period is twice a year. The radio data are binned over 50 days, which is almost equivalent to the period of 53.0--53.2 d.
In addition to binning, the known periodic 53.0--53.2 d flux variation in the radio data is removed in advance, assuming sinusoidal flux variation, because we focus on the comparison of the longer time-scale variability.
The left panel of Figure~\ref{fig:CMP} shows that
both of the flux changes at the W1 and W2 bands are qualitatively consistent with long-scale radio time variability
such as stochastic variability with certain peaks and valleys and almost simultaneous flux variation in the year-scale plot.
A time delay between the MIR and radio bands
is not seen clearly in this long-scale focused plot.

\begin{figure*}
\begin{center}
\includegraphics[width=0.5\linewidth]{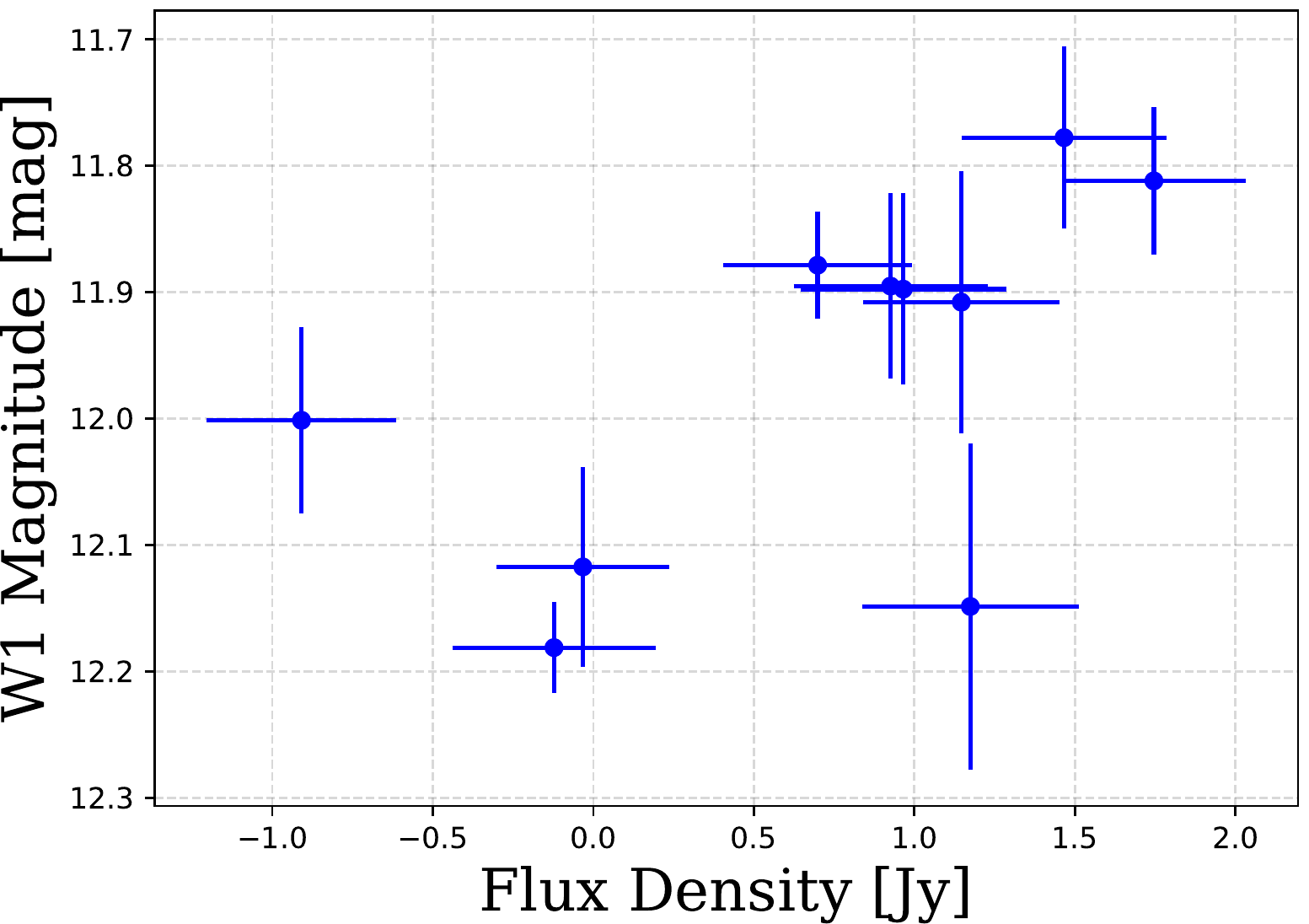}~
\includegraphics[width=0.5\linewidth]{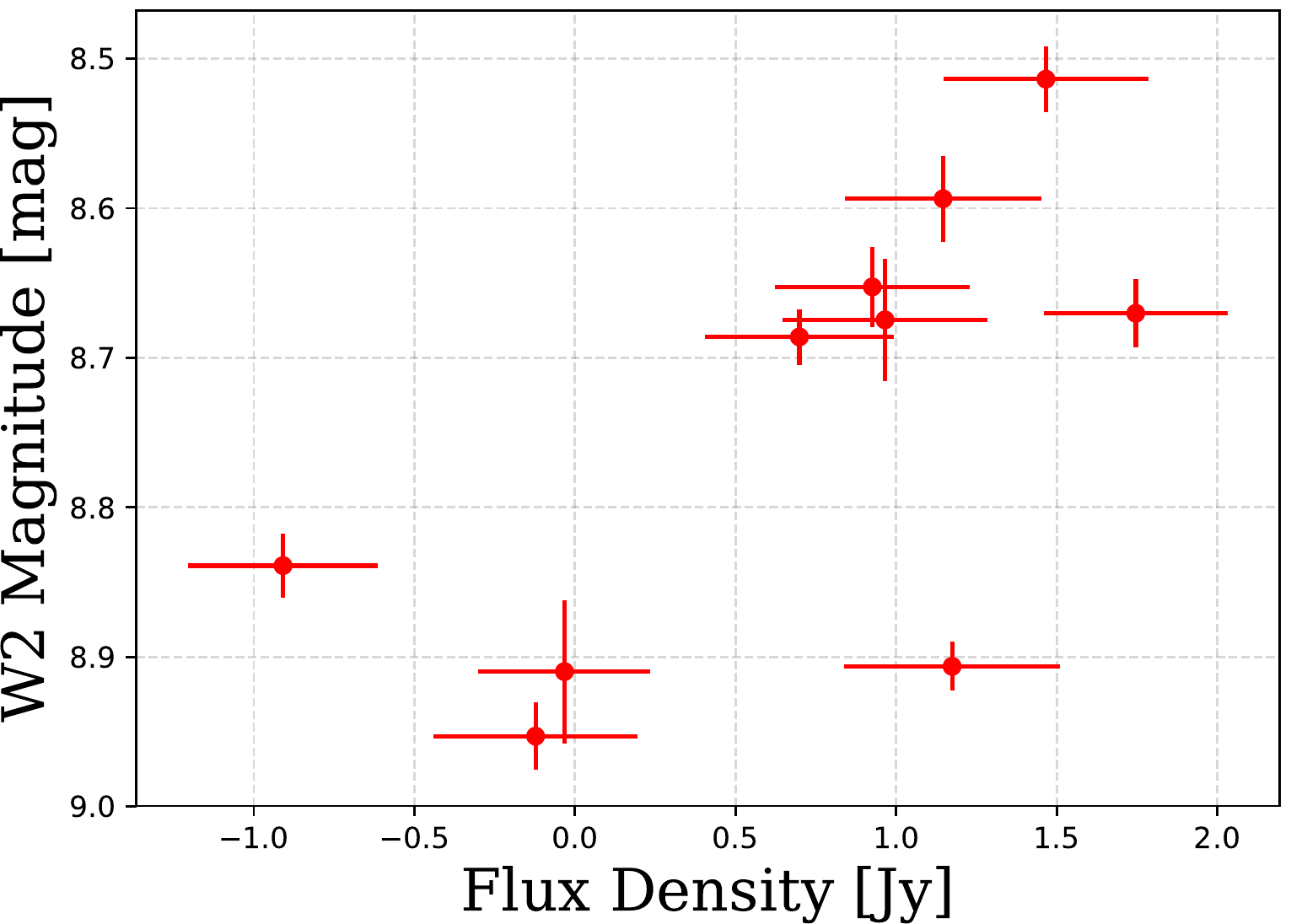}
\caption{
\textit{Left}: Flux correlation plot between W1 band and 6.7~GHz maser emission. It should be noted that the data were subtracted the 53.0-53.2 d known periodic variation, and certain data points indicated negative values.
\textit{Right}: The same plot as the left panel but between W2 band and 6.7~GHz maser emission.
}
\label{fig:FtFcol}
\end{center}
\end{figure*}

We also investigate a more shorter, day-long time variability and the correlation between the MIR and radio data.
We cross-matched the closest 6.7~GHz maser fluxes within 5~days from the each epoch of the NEOWISE observations, and labeled the obtained maser fluxes as ``matched data''.
The results of the matching are presented in the right panel of Figure~\ref{fig:CMP}.
The plot indicates that the flux variations in the MIR and maser data seem to be well correlated without a large time delay.

To check the correlation quantitatively, we also developed MIR and radio intensity matching data plots, as indicated in Figure~\ref{fig:FtFcol}.
Both the W1 and W2 band fluxes indicate a good correlation with the maser flux densities.
The correlation coefficients in the W1 and W2 bands are $+0.59$ and $+0.66$, respectively.
There are certain outlier data points in both the W1 and W2 bands around a flux density of approximately $(F_\nu, W1/W2 [\mathrm{mag}])=(-1.0, 12.0/8.83)$ and $(F_\nu, W1/W2 [\mathrm{mag}])=(1.2, 12.15/8.9)$, which correspond to the first and tenth data points in the right panel of Figure~\ref{fig:CMP}. 
This might be caused by an imperfect fitting and subtraction of the periodic variability or the systematic noise of radio data. 
The correlation coefficients were derived by including these data, which make the coefficients slightly worse.

These results indicate that the MIR and maser emissions are correlated in both the long, year-scale and short, daily-scale variations.
In addition, a clear time delay is not indicated in either timescale.
This indicates that the MIR radiation pumps up 6.7 GHz methanol masers \citep{Cragg05}.

\subsection{Color Variation in the MIR}

\begin{figure*}
\begin{center}
\includegraphics[width=0.33\linewidth]{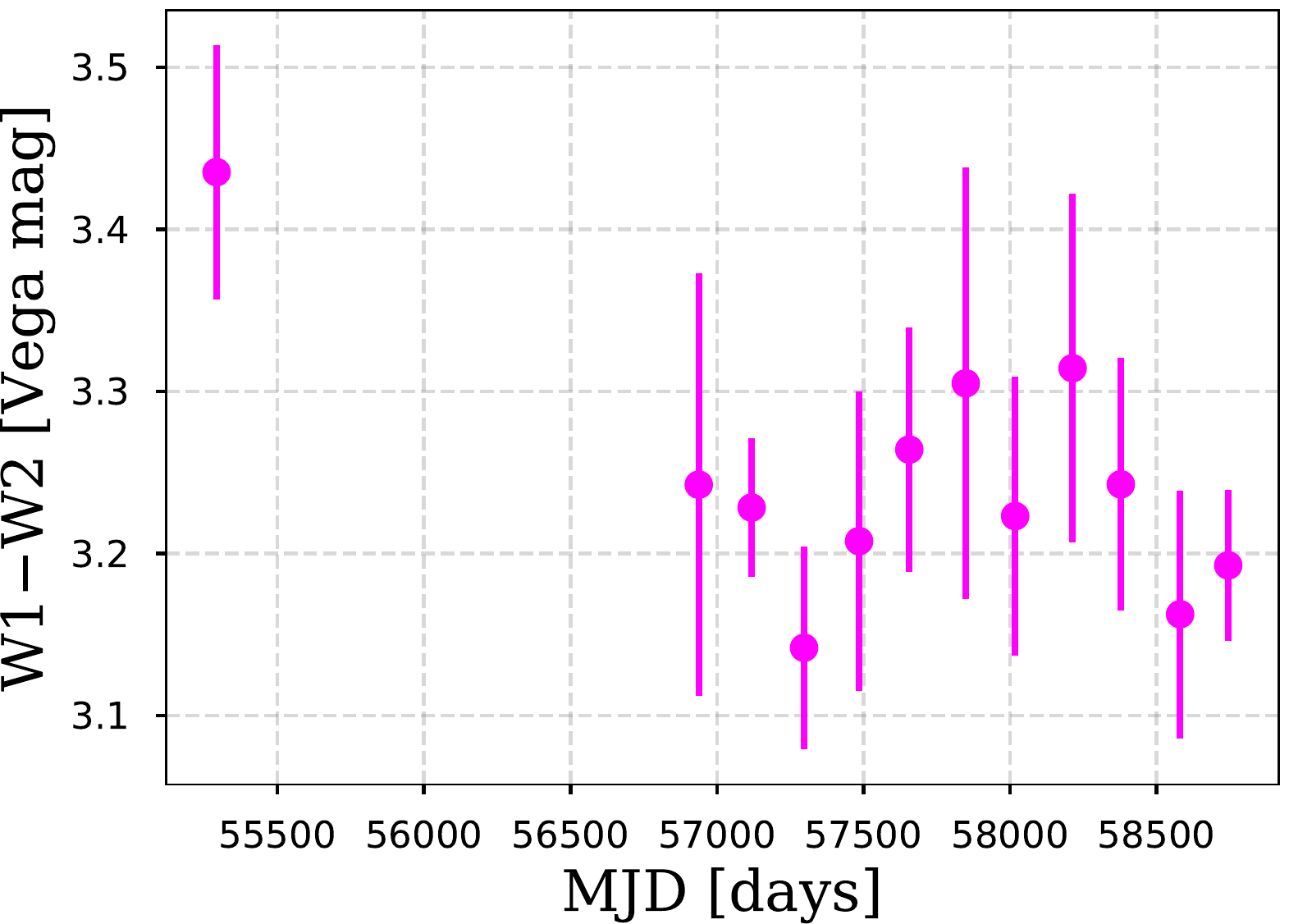}~
\includegraphics[width=0.33\linewidth]{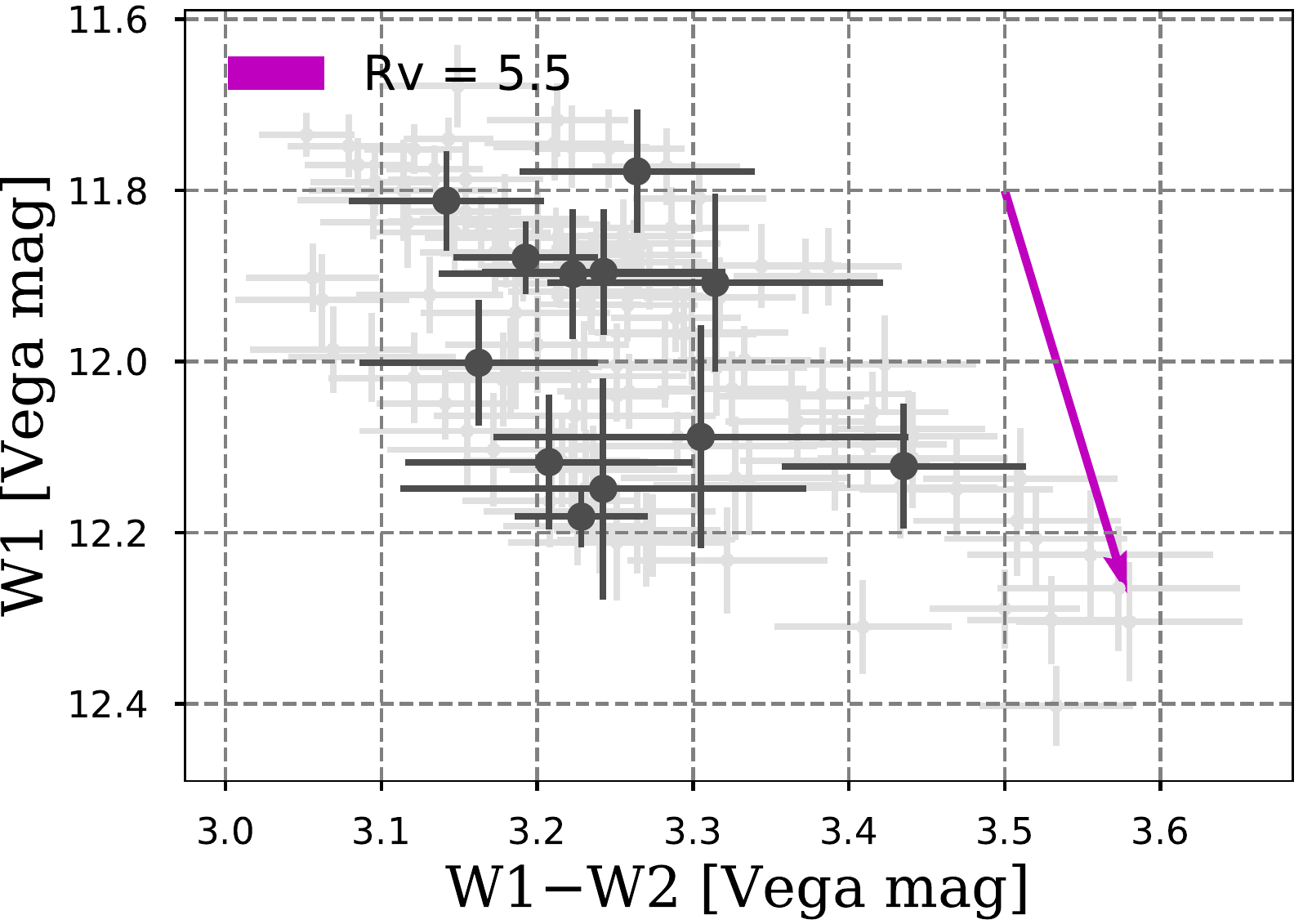}~
\includegraphics[width=0.33\linewidth]{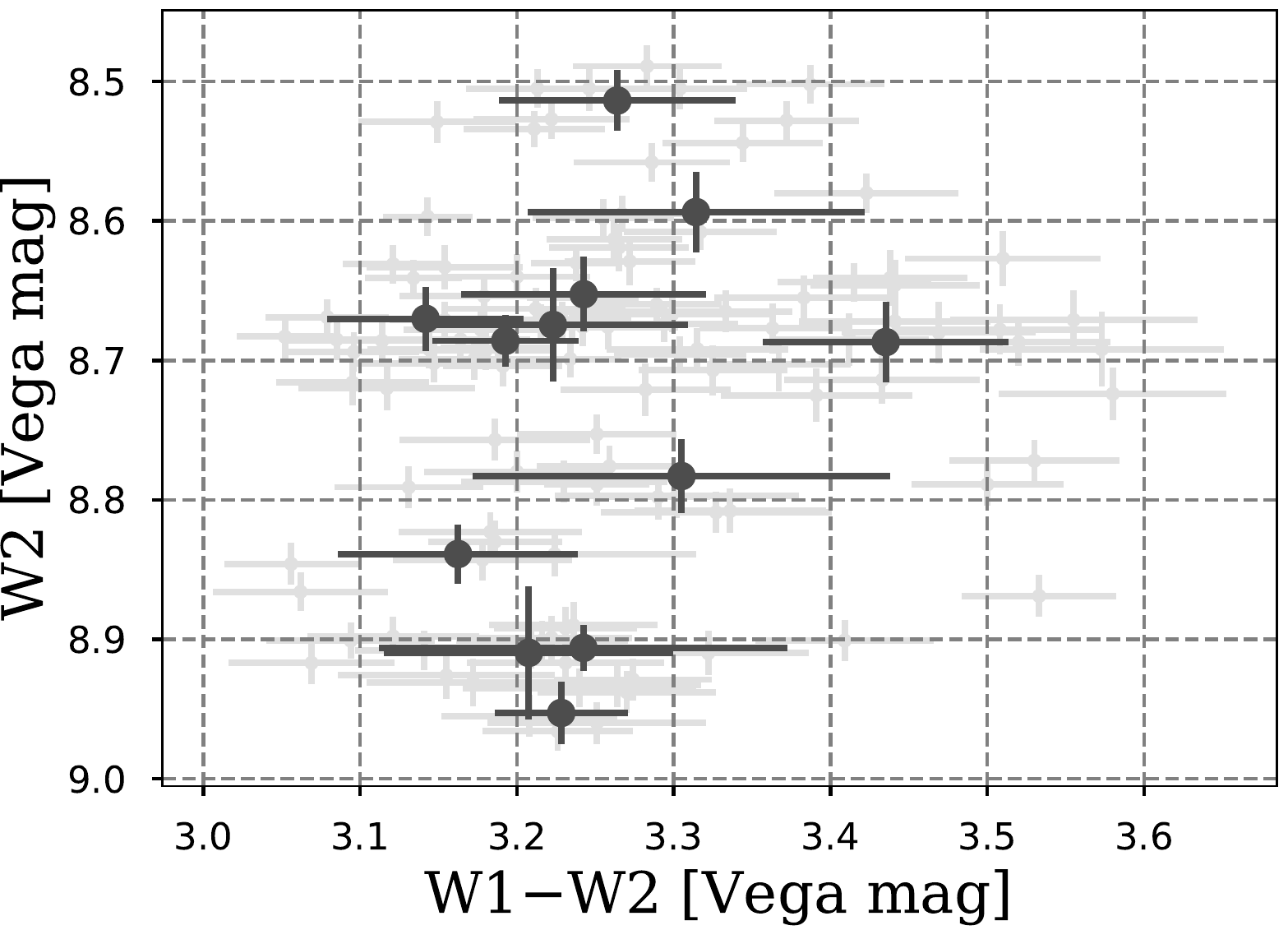}\\
\caption{
\textit{Left}: Light curve of G036.70 in the WISE W1$-$W2 data.
\textit{Middle}: Color magnitude diagram (CMD)
of W1 and W1$-$W2 for G036.70.
Single exposures with their associated error values are
indicated in gray color, and the average value for each
epoch is indicated in black color with error bars.
Magenta vector represents extinction vectors at $R_\mathrm{V}=5.5$.
\textit{Right}: CMD of W2 and W1$-$W2. }
\label{fig:CMD}
\end{center}
\end{figure*}

WISE color variability is a good proxy for deciphering two possible origins of variability: extinction and intrinsic flux variability \citep[for example,][]{uch19}.
The left panel of Figure~\ref{fig:CMD} presents a light curve of W1$-$W2 color throughout the whole WISE data with a 1-$\sigma$ error bar.
The resultant plot indicates that there is no significant color variability in G036.70 NEOWISE data, at least between the W1 and W2 bands.

The middle and right panel of Figure~\ref{fig:CMD} shows the two color-magnitude diagrams (CMDs) of W1/W2 and W1$-$W2 for G036.70 during the preceding 10~yr period.
The gray cross represents the data for a single exposure, whereas the black cross represents the averaged data binned for each epoch.
It should be noted that the correlation seen in the gray cross of the middle panel is
made by the artificial trend because both of the W1 vs W1$-$W2 data contains the W1 band magnitude and therefore the noise fluctuation by W1 produces an artificial $y = -x$ linear sequence.
Considering that we are interested in the year-scale long-term variability, we focus here on the trend of the binned data.
The middle panel indicates that the averaged data do not display a clear color-dependent trend. 
To determine the details, we also developed the right, W2 vs W1$-$W2, plot.
The resultant plot indicates that the W1$-$W2 color does not change significantly, and its behavior seems to be stochastic, whereas the W2 flux changes as much as 0.4 mag.
These results indicate that the MIR color 
of W1$-$W2 does not change significantly with the flux variation in G036.70. 
We considered that this flux variation without color change would be intrinsic while we cannot totally rule out the situation that this trend occurred by line-of-sight extinction.

\section{Discussion}

\subsection{Stellar Physical Parameters of G036.70}
To discuss the flux variability of G036.70 in detail, we first investigated the stellar physical parameters through stellar model fitting using the IR spectral energy distribution (SED).
The IR photometric data consist of archived data from \textit{Spitzer}, \textit{WISE}, and \textit{Herschel} \citep{glimpse09,MIPS15,wise12,HiGAL17}  by cross-matching the coordinates of the maser emission of G036.70 with a search radius of 2.5 arcsec. The obtained IR SED spans the wavelengths from $\sim3$ to $350$~$\mu$m, as summarized in Table~\ref{tab:SED}, which covers the dust emission peak at 100~$\mu$m.

We applied the SED fitting tool to the IR SED provided by \cite{Robitaille07}. It contains various combinations of stellar and accretion disk parameters, such as extinction, total bolometric luminosity, stellar mass, effective temperature, and disk mass.

\begin{figure}
\begin{center}
\includegraphics[width=1.0\linewidth]{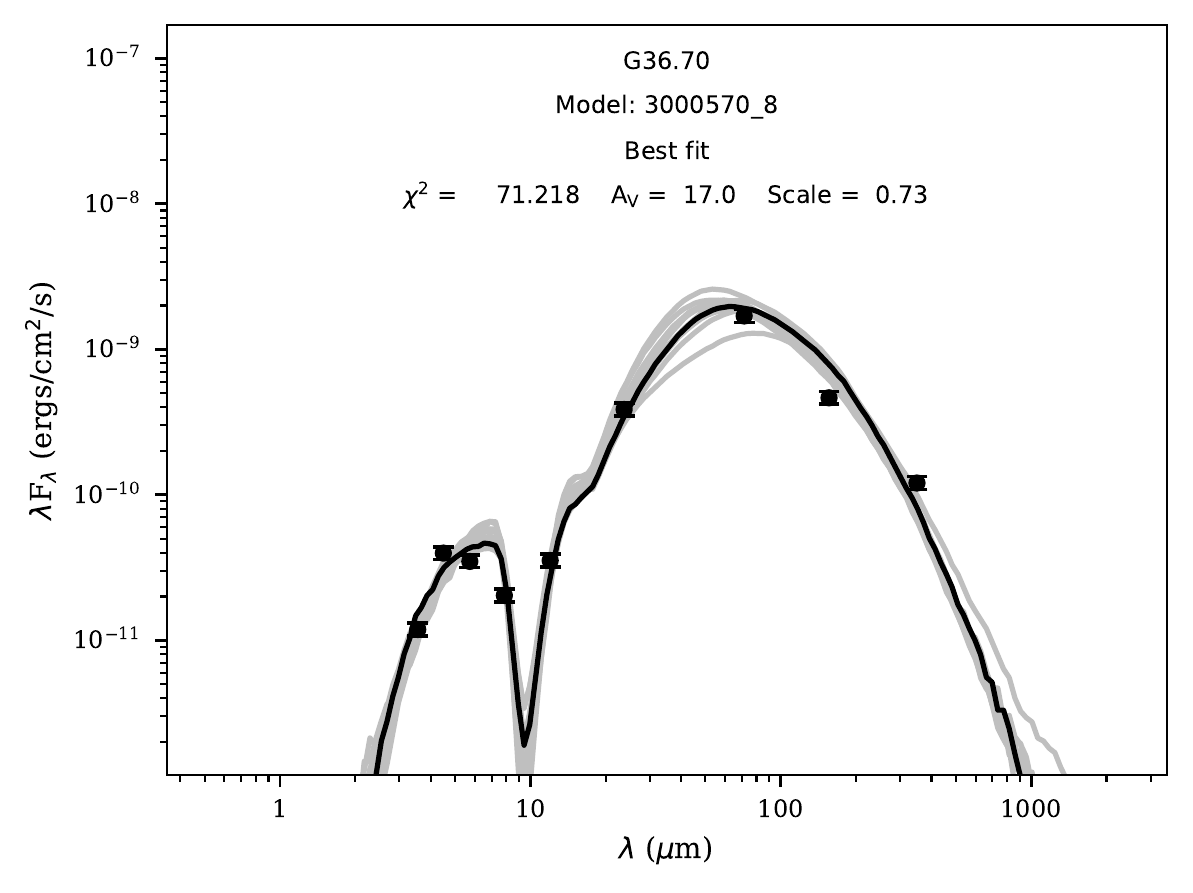}~
\caption{Obtained spectral energy distribution (SED) for G036.70 and 
the best-fitting result. The black points with error bars 
correspond to the obtained infrared (IR) photometries from
\textit{Spitzer}, \textit{Herschel}/PACS, and SPIRE.
The best-fitting SED is represented by a solid black line.
}
\label{fig:SEDfit}
\end{center}
\end{figure}

Figure~\ref{fig:SEDfit} depicts the model SED (gray-colored area) obtained from the SED fitting provided by \citet{Robitaille07} as well as the obtained data points (black points). The SED model well reproduces the peak emission at $\sim100$~$\mu$m and the expected deep silicate absorption at $\sim10$~$\mu$m as well as the strong dust extinction at $\lambda<2$~$\mu$m.
The obtained best-fit stellar parameters are tabulated in Table~\ref{tab:G36para}, assuming a distance range of 4$-$7 kpc from the gas kinematic distance measurement \citep{Szymczak07,Pandian07}.
The obtained parameters indicate that G036.70 is essentially a massive and obscured YSO with a mass of 10 $M_{\odot}$.
The derived effective temperature of less than $10^4$ K is well consistent with previous observations of the absence of HII regions \citep{Svoboda16}.
Note that the SED fitting provides a crude estimate of stellar parameters because all the measurements used to fit SED originates from different measurement epochs, and the most parameters listed in Table~\ref{tab:G36para} could be treated with care.

In addition to the SED profile, the \textit{Spitzer} archival color image of G036.70 shows ``green'' and ``extended'' feature, a similar one of the extended green objects \citep{Cyganowski08}. This suggests that G036.70 is an active accreting object with powerful outflows and has a nearly edge-on disk, which is also indicated by the deep silicate feature in the IR SED.

\subsection{Possible Origins of MIR and maser Variability in G036.70}

The results presented in Section~\ref{sec:results} indicate that G036.70 has i) year-scale stochastic MIR and maser flux variability, ii) no significant color variability in the MIR, and iii) a clear correlation with the 6.7~GHz maser intensity for both year-long and day-long timescales.
In this section, we discuss the possible origins of the MIR variability of G036.70.
It should be noted that we will not discuss the known 53.0-53.2 d periodic variability in masers because the time cadence of the WISE MIR data is not enough to compare with this periodic variability. Future high-cadence MIR monitoring observations will enable us to discuss the possible origins of the periodic variability with MIR short timescale variability.

The year-scale flux variability suggests non-periodic events such as changes in the mass accretion rate or line-of-sight extinction owing to non-axisymmetric dust density distribution in the rotating accretion disk. 
In contrast, stochastic MIR flux variability rules out the stellar and/or planet occultation origins that indicate no MIR color variations.
Therefore, the possible scenario is the change in mass accretion rate or the line-of-sight extinction owing to non-axisymmetric dust density distribution of the edge-on like accretion disk, with no significant color variation. 
This may occur because the color change caused by these scenarios is considered to be not as large as $\Delta$W1$-$W2 $< 0.5$~mag \citep{WLJ14,CoG17}. 
We cannot rule out the origin of the dust extinction because of the large scatter depicted in Figure~\ref{fig:CMD}.
Thus, the trend is still consistent with the interstellar dust extinction vector.

In addition, the promising origin must describe the observed MIR-maser variability correlation in our study.
Variability caused by change in the accretion rate triggers changes in both the stellar luminosity and disk surface temperature distribution, and thus the disk infrared emission intensity distribution, such phenomenon was reported in S255-NIRS3 accretion burst \citep{CoG17}.
Therefore, the change in accretion rate can explain the MIR-maser variability correlation because the class II methanol maser is considered to be pumped up by disk IR emission \citep{Cragg05}.
We can examine this scenario in the future more precisely through spectroscopic monitoring of the NIR and MIR bands covering accretion tracer emission lines, such as hydrogen recombination lines.
For variability caused by changes in line-of-sight extinction, some of the major scenarios cannot describe the observed MIR-maser correlation, such as the occulation of the flaring outer disk or infalling gas and dust, because such events do not affect the disk infrared emission intensity.
One possible origin is the puff-up inner rim \citep{Dullemond01}.
The non-axisymmetric dust density distribution of the inner rim can change the line-of-sight extinction through occultation, and it can also change the disk infrared emission intensity by creating a shaded area on the accretion disk.
In such a case, it is essential to monitor the MIR emission at longer wavelengths to examine whether color variability occurs, which was previously reported in the lower-mass puff-up inner rim YSO system.

\subsection{Importance of correlation between MIR and maser variability detection in G036.70}

As we discussed above, MIR-maser variability correlation in MYSOs was first suggested in S255-NIRS3 during the accretion burst phase \citep{CoG17}. 
The authors reported the appearance of accretion tracers during the changes of SED.
Soon after this report, \citet{Stecklum18} and \citet{Olech20} showed the 
periodic MIR-maser variability correlation in the intermediate mass YSO, G107.298+5.639.
Therefore, our finding on the correlation between the MIR--maser variabilities in G036.70 is the second case for the MYSOs, and the first case in the non-burst phase MYSOs.
This detection further confirms the proposed scenario that the class II methanol maser in MYSOs is pumped up by disk IR emission \citep{Cragg05}, not by outflow related shock.

Our results also indicate that the time delay between MIR and maser flux variation is less than a few days. 
This short time delay suggests a scenario that variations in the mass accretion rate of a MYSO can trigger the accretion luminosity change and hence the MIR flux change. It immediately affects the temperature distribution of the accretion disk, which results in the trigger of the maser flux change.
The scenario hence suggests the situation that the outflow cavity is a real ``cavity'', which does not contain MIR optically thick gas or dust causing a significant time delay.
Further precise measurements of time delay and more multi-wavelengths monitoring of MYSOs will reveal characteristics and mechanisms of correlation between MIR and maser flux variability in MYSOs.

\begin{deluxetable}{lccc}
\tablecaption{
IR SED of G036.70 used for IR SED fitting\label{tab:SED}}
\tablehead{
\colhead{filter} &
\colhead{$\lambda$} &
\colhead{$f_\nu$} & 
\colhead{Aperture}\\
-- & [$\mu$m] & [Jy] & [arcsec]
}
\startdata
\hline
IRAC3.6 & 3.6 & 0.0141 & 2.4 \\
IRAC4.5 & 4.5 & 0.0598 & 2.40 \\
IRAC5.8 & 5.8 & 0.0669 & 2.40 \\
IRAC8.0 & 8.0 & 0.0536 & 2.40 \\
WISE12 & 12.0 & 0.1420 & 8.25 \\
MIPS24 & 24.0 & 3.0600 & 6.50 \\
PACS70 & 70.0 & 40.5460	& 6.00 \\
PACS160 & 160.0 & 24.2280 & 12.00 \\
SPIRE350 & 350.0 & 14.1190 & 24.00 \\
\hline
\enddata
\tablecomments{Columns:
The four columns are defined as follows.
Filter: Filter name used for SEDfitter.
$\lambda$: Central wavelength with unit $\mu$m.
$f_\nu$: Flux density with unit Jy.
Aperture: Aperture size of each photometry.
}
\end{deluxetable}

\begin{deluxetable}{lcc}
\tablecaption{
The main parameters for the best-fitting models of G036.70\label{tab:G36para}}
\tabletypesize{\small}
\tablecolumns{2}
\tablewidth{5pt}
\tablehead{
\colhead{Col.} &
\colhead{Parameter} &
\colhead{G036.70}
}
\startdata
\multicolumn{3}{c}{General}\\
\hline
(1) & $A_\mathrm{V}$/mag & $ 25.9_{-19}^{+28} $\\
(2) & $L_\mathrm{tot}/L_{\odot}$ & $2.9_{-1.0}^{+0.97} \times10^{3}$ \\
\hline
\multicolumn{3}{c}{Central source}\\
\hline
(3) & $M_{\star}/M_{\odot}$ & $10.0_{-2.0}^{+0.8}$ \\
(4) & $R_{\star}/R_{\odot}$ & $35.0_{-18}^{+18}$ \\
(5) & $T_{\star}/{\rm K}$ & $7.4_{-2.8}^{+2.7}\times10^{3}$\\
\hline
\multicolumn{3}{c}{Disk component}\\
\hline
(6) & $M_{\rm disk}/M_{\odot}$ & $1.6_{-1.1}^{+8.9}\times10^{-2}$ \\
(7) & $R_\mathrm{disk,in}/\mathrm{au}$ & $4.3_{-1.3}^{+0.7}$ \\
(8) & $R_\mathrm{disk, out}/\mathrm{au}$ & $1.8_{-1.3}^{+1.3}\times10^2$ \\
\enddata
\tablecomments{Columns: (1) Foreground dust extinction.
(2) Total bolometric luminosity of G~036.70.
(3) Central stellar mass.
(4) Central stellar radius.
(5) Central stellar temperature.
(6) Disk mass.
(7) Disk inner radius.
(8) Disk outer radius.}
\end{deluxetable}

\section{Conclusions}

We identified a stochastic colorless yearly variability of G036.70 in both the WISE W1 (3.4~$\mu$m) and W2 (4.6~$\mu$m) bands using the
ALLWISE and NEOWISE archival data covering a 10-yr-long time period with 6-month cadence.
We also found the flux variability correlation between the obtained WISE W1/W2 bands and the 6.7-GHz class II methanol maser fluxes using the high-cadence monitoring data obtained using the Hitachi 32-m radio telescope: the Ibaraki 6.7-GHz Methanol Maser Monitor (iMet) program \citep{Yonekura16,Sugiyama19}.
The correlation can be seen in both the year-scale light curve and the much shorter daily timescale.
This MIR-maser variability correlation was reported for the first time in non-burst MYSOs.
Our results support the scenario that a class II methanol maser is pumped up by infrared emission from accreting disks of MYSOs.

We also discussed the possible origins of MIR and maser variability.
Considering the observed phenomena, such as yearly stochastic flux variation, almost colorless variability, and MIR-maser correlation, 
a change in the mass accretion rate or line-of-sight extinction caused by crossing puffed-up inner rim may describe the situation.
Monitoring observations through spectroscopy can help further investigate the origin of variability in G036.70.

\acknowledgments

We appreciate the support of Sizuka Nakajima for collecting essential data.
We would like to thank Yu Saito, Naoko Furukawa, Hiroshi Akitaya, and the students at Ibaraki University who supported the maser observations.
This work is partially supported by the Inter-university collaborative project "Japanese VLBI Network (JVN)" of NAOJ.
This study benefited from financial support from the Japan Society for the Promotion of Science (JSPS) KAKENHI program (18K13584, KI; 21H01120 and 21H00032, YY; and 19K03921, KS),
and the Japan Science and Technology Agency (JST) grant ``Building of Consortia for the Development of Human Resources in Science and Technology'' (KI).
Finally, this study made use of data products from the Near-Earth Object Wide-field Infrared Survey Explorer (NEOWISE), a project of the Jet Propulsion Laboratory (JPL)/California Institute of Technology. NEOWISE is funded by the National Aeronautics and Space Administration (NASA).

\facilities{\textit{WISE}}

\software{astropy \citep{Astropy13}, Matplotlib \citep{hun07}, Pandas \citep{mck10}}


\bibliographystyle{apj}
\bibliography{apj-jour,ref.bib}

\end{document}